\documentclass[12pt]{article}
\newcommand{\IR}{\mbox{$I\!\!R$}}
\newcommand{\IN}{\mbox{$I\!\!N$}}
\newcommand{\out}{{{\cal O}}}
\newcommand{\complex}{\C}
\newcommand{\D}{{\cal D}}
\newcommand{\C}{{\cal C}}
\newcommand{\M}{{\cal M}}
\newcommand{\Pos}{{\mathit{Pos}}}
\newcommand{\Neg}{{\mathit{Neg}}}

\newtheorem{theorem}{Theorem}[section]
\newtheorem{example}[theorem]{Example}
\newcommand{\qed}{\vrule height7pt width4pt depth1pt \vspace{0.1in}}
\newcommand{\Prime}{\mathit{Prime}}
\newcommand{\commentout}[1]{}
\newcommand{\citeyear}{\cite}

\begin{document}
\title{Decision Theory with Resource-Bounded Agents}
\author{
Joseph Y. Halpern\thanks{Supported in part by NSF grants IIS-0812045, 
IIS-0911036, and CCF-1214844, AFOSR grants 
FA9550-08-1-0438 and FA9550-09-1-0266, and ARO grant W911NF-09-1-0281.} \\
Cornell University\\
halpern@cs.cornell.edu
\and Rafael Pass\thanks{Supported in part by a 
Alfred P. Sloan Fellowship, a Microsoft New Faculty Fellowship, NSF
CAREER Award CCF-0746990, 
NSF grant CCF-1214844,
AFOSR YIP Award FA9550-10-1-0093, and DARPA
and AFRL under contract FA8750-11-2- 0211. 
The views and conclusions contained in this document are those of the
authors and should not be interpreted as representing the official
policies, either expressed or implied, of the Defense Advanced Research
Projects Agency or the US government.}\\
Cornell University\\
rafael@cs.cornell.edu
\and Lior Seeman\thanks{Supported in part by NSF grants IIS-0812045
 and CCF-1214844, and AFOSR grants 
FA9550-08-1-0438 and FA9550-09-1-0266.} \\
Cornell University\\
lseeman@cs.cornell.edu}
\maketitle

\begin{abstract}
There have been two major lines of research aimed at capturing
resource-bounded players in game theory.
The first, initiated by Rubinstein \citeyear{Rub85}, 
charges an agent for doing costly computation;
the second, initiated by Neyman \citeyear{Ney85}, 
does not charge for computation, but limits the computation that agents
can do, 
typically by modeling 
agents as finite automata.  We review recent work on applying
both approaches in the context of decision theory.  
For the first approach, we take the objects of choice in a decision
problem to be Turing machines, 
and charge players for the ``complexity'' of the Turing machine
chosen (e.g., its 
running time).  This approach can be used to explain
well-known phenomena like \emph{first-impression-matters biases} (i.e.,
people tend to put more weight on evidence they hear early on)
and 
\emph{belief polarization} (two people with different prior beliefs,
hearing the same evidence, can end up with diametrically opposed
conclusions) as the outcomes of quite rational decisions. 
For the second approach, we model people as finite
automata, and provide a simple algorithm that, on a problem that
captures a number of settings of interest, provably performs
optimally as the number of states in  
the automaton increases.
\end{abstract}

\section{Introduction}
The standard approach to decision making, going
back to Savage \citeyear{Savage}, suggests that an agent should maximize
expected utility.  But computing the relevant
probabilities might be difficult, as might computing the relevant
utilities.  And even in cases where the probabilities and utilities are
not hard to compute, finding the action that maximizes expected utilities
can be difficult.  In this paper, we consider approaches to decision
making that explicitly take computation into account.

The idea of taking computation into account goes back at least to the
work of Good \citeyear{Good52} and Simon \citeyear{Simon55}.  It has
been a prominent theme in the AI literature (see, e.g.,
\cite{Hor87,RS95}).  Our work has been inspired by 
two major lines of research aimed at capturing
resource-bounded 
(i.e. computationally-bounded) agents in the game theory literature.
The first, initiated by Rubinstein \citeyear{Rub85}, 
charges an agent for doing costly computation;
the second, initiated by Neyman \citeyear{Ney85}, 
does not charge for computation, but limits the computation that agents
can do, 
typically by modeling
agents as finite automata.  
We review recent work on applying
both approaches in the context of decision theory.  

We consider the first approach in the context of the framework of
\cite{HP10a}, which in turn is a specialization of the framework of
\cite{HP08} to the case of a single agent.  The idea is the following:
We assume that the agent can be viewed as choosing an algorithm
(i.e., a Turing machine); with each Turing machine (TM) $M$ and input, we
associate its \emph{complexity}.   The complexity can represent, for
example, the running time of $M$ on that input, the space used,
the complexity of $M$ (e.g., how many states it has), or the difficulty of
finding $M$ (some algorithms are easier to think of than others).  We
deliberately keep the complexity function abstract, to allow for the
possibility of representing a number of different intuitions.  The agent's
utility can then depend, not just on the payoff, but on the complexity.
Thus, we can ``charge'' for the complexity of computation.

Although this approach seems quite straightforward, we show that it can
be used to explain well-known phenomena like
\emph{first-impression-matters biases} (i.e., people tend to put more
weight on evidence they hear early on) and \emph{belief polarization}
(people with different prior beliefs, hearing the same evidence, can end
up with diametrically opposed conclusions).

We then consider the second approach: modeling agents as finite
automata.  
Finite automata are a well-known basic model of computation.
An automaton receives a sequence of inputs from the environment, and
changes state in response to the input.  We capture the fact that an
agent is resource bounded by limiting the number of states in the
automaton. 

In the context of decision theory, this approach was first
used by Wilson \citeyear{W02}.  She examined a decision problem where
an agent needs to make a single decision, whose payoff depends on the
state of nature (which does not change over time).  Nature is in one of
two possible states, $G$ (good) and $B$ (bad).  The agent gets signals,
which are correlated with the true state, until the game ends, 
which happens at each step with probability $\eta > 0$.
At this point, the agent must make a decision.  For each $N$, 
Wilson characterizes an $N$-state optimal finite automaton for making a
decision in this  setting, under the assumption that $\eta$
is small (so that the agent gets information for many rounds).  (See
Section~\ref{sec:aaai12} for further details.)  
She then uses this characterization to argue that an optimal $N$-state
automaton also exhibits behavior such as  belief polarization
(again, see Section~\ref{sec:aaai12}).
Thus, some observed human behavior can be explained by
viewing people as resource-bounded, but rationally making the best use
of their resources (in this case, the limited number of states).

Wilson's model assumes that nature is static.  But in many important
problems, ranging from investing in the stock market to deciding which
route to take when driving to work, the world is dynamic.
Moreover, people do not make decisions just once, but must make them often.
For example, when investing in stock markets, people get signals about
the market, and need to decide after each signal whether to invest more
money, take out money that they have already invested, or to stick
with their current position.  In recent work \cite{HPS12}, we consider
the problem of constructing an optimal $N$-state automaton for this
setting.    We construct a family of quite simple automata, indexed by
$N$, the number of states, and a few other parameters. 
We show
that as $N$ grows large, this family approaches optimal behavior.
More precisely, for all $\epsilon$, there is a sufficiently
large $N$ and member of this family with $N$ states whose expected
payoff is within $\epsilon$ of optimal, provided the probability of a
state transition is sufficiently small.
More importantly, the members of this family reproduce observed human behavior in a series of tasks
conducted by Erev, Ert, and Roth \citeyear{erev2010entry} (see
Section~\ref{sec:aaai12} for further discussion%
). 
Again, these results emphasize the fact that some observed human
behavior can be explained by viewing people as rational,
resource-bounded agents.  


\section{Charging for the complexity of computation}

In this section we review and discuss the approach that we used in
\cite{HP10a} to take cost of computation into account in 
decision theory.  Much of the material below is taken from \cite{HP10a},
which the reader is encouraged to consult for further details and
intuition.  

The framework that we use is 
essentially a single-agent version of what we called in \cite{HP10,HP08}
\emph{Bayesian machine games}.  
In a standard Bayesian game, each player
has a \emph{type} in some set $T$, and makes a single move.  
Player $i$'s type can be viewed as describing $i$'s
initial information; some facts that $i$ knows about the world.  
We assume that an agent's move consists of choosing a Turing machine.  
As we said in the introduction, associated with each
Turing machine and type is its complexity.   Given as input a type, the
Turing machine outputs an action.  The utility of a player depends on
the type \emph{profile} (i.e., the types of all the players), the
action profile, and the complexity profile (that is, each player's
complexity).  


\begin{example}
\label{example:ex1}
{\rm
Suppose that an agent is given an input $n$, and is asked
whether it is prime.  The agent gets a payoff of \$1,000 if he gives the
correct answer, and loses \$1,000 if he gives the wrong answer.  However,
he also has the option of playing safe, and saying ``pass'', in which
case he gets a payoff of \$1.  Clearly, many agents would say ``pass'' on
all but simple inputs, where the answer is obvious,
although what counts as a ``simple'' input may depend on the agent.%
\footnote{While primality testing is now known to be in polynomial time
\cite{AKS02}, and there are computationally-efficient randomized
algorithms that that give the correct answer with extremely high
probability \cite{Rabin80,SolovayS77}, we can assume that the agent has no
access to a computer.}  
The agent's type can then be taken to be the input.  The set $T$ of
possible types could be, for example, all integers, or all integers 
that can be written in binary using at most 40 digits (i.e., a number
that is less than $2^{40}$).    The agent can
choose among a set of TMs, all of which output either ``prime'', ``not
prime'' or ``pass'' (which can be encoded as 0, 1, and 2).  One natural
choice for the complexity of a pair $(M,n)$ consisting of TM $M$
and an input $n$ is the running time of $M$ on input $n$.}

{\rm 
Since the agent's utility takes the complexity into account, this would
justify the agent using a ``quick-and-dirty'' algorithm which is typically
right to compute whether the input is prime; if
the algorithm does not return an answer in a relatively small amount of
time, the agent can just ``pass''.  The agent might prefer using such a TM
rather than one that gives the correct answer, but takes a long time
doing so.  We can capture this tradeoff in the agent's utility function.
}
\qed
\end{example}

There are some issues that we need to deal with in order to finish
modeling Example~\ref{example:ex1} in our framework.  Note that if
the agent chooses the TM after being given the number $n$, then one of two
TMs is clearly optimal: if $n$ is in fact prime, then the TM which just
says ``prime'' and halts; if $n$ is not prime, then the TM which says
``not prime'' and halts is clearly optimal.  The problem, of course, is
that the agent does not know $n$.  One approach to dealing with this
problem is to assume that the agent chooses the TM before knowing $n$; this
was the approach implicitly taken in \cite{HP08}.  But there is 
a second 
problem:  we have implicitly assumed that the agent knows what the
complexity of each pair $(M,n)$ is; otherwise, the agent could not maximize
expected utility.  Of course, in practice, an agent will not know how long
it will take a Turing machine to compute an answer on a given input.  

We deal with both of these problems by adding one more parameter to the
model: a state of nature.  In Example~\ref{example:ex1}, we allow the
TM's running time and whether $n$ is prime to depend on the state; this
means that the ``goodness'' (i.e., accuracy) of the TM can depend on the
state.  We model the agent's beliefs about the running time and accuracy
of the TM using a probability distribution on these states.%
\footnote{This means that the states are what philosophers have called 
``impossible'' possible worlds \cite{Hi2,Rant}; for example, we allow
states where $n$ is taken to be prime even when it is not, or where a TM
halts after $M$ steps even if it doesn't.  We need such states, which are
inconsistent with the laws of mathematics, to model a resource-bounded
agent's possibly mistaken beliefs.}   

We capture these ideas formally by taking
a \emph{computational decision problem with types} to be a tuple
$\D = (S,T,A,\Pr,\M,\complex, \out, u)$.  We explain each of these
components in turn.
The first four components are fairly standard.
$S$ is a state space, $T$ is a set of types,
$A$ is a set of actions, and $\Pr$ is a probability distribution on $S
\times T$ (there may be correlation between states and types).
In a standard decision-theoretic setting, there is a probability on
states (not on states and types), and the utility function associates a
utility with each (state, action) pair (intuitively, $u(s,a)$ is the
utility of performing action $a$ in state $s$).  
Here things are more
complicated because we want the utility to also depend on the TM chosen.
This is where the remaining components of the tuple come in.

%
The fifth component of the tuple, $\complex$, is the \emph{complexity
function}. 
Formally, $\complex: {\bf M} \times S \times T \rightarrow \IN$, so
$\complex(M,s,t)$ is the complexity of running TM $M$ on input $t$ in
state $s$.  The complexity can be, for example, the running time of $M$
on input $t$, the space used by $M$ on input $t$, the number of states
in $M$ (this is the measure of complexity used by Rubinstein
\citeyear{Rub85}), and so on.  The sixth component of the tuple, $\out$,
is the 
{\em output function}.  It captures the agent's uncertainty about the
TM's output.  Formally, $\out : {\bf M} \times S \times T   
\rightarrow \IN$;  
$\out(M,s,t)$ is the output of $M$ on input $t$ in state $s$.
Finally, an agent's utility depends on the state $s$, his type $t$, and the
action $\out(M,s,t)$, as is standard, and the complexity.
Since we describe the complexity by a natural number, we take the
utility function $u$ maps $S \times T \times A \times \IN$ to $\IR$ (the
reals).
Thus, the expected utility of choosing a TM $M$ in the decision problem
$\D$ is
$\sum_{s \in S, t \in T} \Pr(s,t) u(s,t,\out(M,s,t),\complex(M,s,t))$.
Note that now the utility function gets the complexity of $M$ as an
argument.  The next example should clarify  the role of all these components.

\medskip
\noindent {\bf Example~\ref{example:ex1}} (cont'd):  We now have the
machinery to formalize Example~\ref{example:ex1}.  We take $T$, the type
space, to consist of all natural numbers
$< 2^{40}$;  the agent must determine
whether the type is prime.  The agent can  choose either 0 (the number
is not prime), 1 (the number is prime), or 2 (pass); 
Thus, $A = \{0,1,2\}$.
Let $\M$ be some set of TMs that can be used to test for primality.  
As suggested above, the state space $S$ is used to capture the agent's
uncertainty about the output of a TM $M$ and the complexity of $M$.
Thus, for example, if the agent believes that the TM will output pass with
probability $2/3$, then the set of states such that $\out(M,s,t) = 2$ has
probability $2/3$.
We take $\complex(M,s,t)$ to be $0$ if $M$ computes the answer within $2^{20}$
steps on input $t$, and 10 otherwise.  
(Think of $2^{20}$ steps as representing a hard deadline.)
If, for example, the agent does not
know the running time of 
a TM
$M$, but ascribes probability $2/3$ to $M$
finishing in less than $2^{20}$ steps on input $t$, then the set of
states $s$ such that $\C(s,t,M) = 0$ has probability $2/3$.
We assume that there is a function $\Prime$ that captures the agent's
uncertainty regarding primality; $\Prime(s,t) = 1$ if $t$ is prime in
state $s$, and 0 otherwise.%
\footnote{Thus, we are allowing ``impossible'' states, where $t$ is
viewed as prime even if it is not.}  
Thus, if the agent believes that TM $M$ gives the right answer with
probability $2/3$, then the set of states $s$ where $\Prime(s,t) =
\out(M,s,t)$ has probability $2/3$.
Finally, let $u(s,t,a,c) = 10 - c$ if $a$ is either 0 or 1, and
this is the correct answer in state $s$ 
(i.e., $\out(M,s,t) = \Prime(s,t)$)
and $u(s,t,2,c) = 1-c$.  
Thus, if the agent is sure that $M$ always gives
the correct output, then $u(s,t,a,c) = 10 - c$ for all states $s$ and $a
\in \{0,1\}$.
%
\qed

\begin{example}
[Biases in information processing]
\label{informationbias:ex} 
{\rm Psychologists have observed many systematic biases in the way that
individuals update their beliefs as new information is received (see
\cite{mrabin98} for a survey). In particular, a
\emph{first-impressions} bias has been observed: individuals put too
much weight on initial signals and less weight on later signals. As they
become more convinced that their beliefs are correct, many individuals
even seem to simply ignore all information once they reach a confidence
threshold. 
Several papers in behavioral economics have focused on
identifying and modeling some of these biases (see, e.g., \cite{mrabin98}
and the references therein, \cite{M98}, and \cite{RS99}). In
particular, Mullainathan \citeyear{M98} makes a potential connection
between memory and biased 
information processing, using a model that makes several explicit
(psychology-based) assumptions on the memory process (e.g., that the
agent's ability to recall a past event depends on how often he has
recalled the event in the past). 
More recently, Wilson \citeyear{W02} demonstrated a similar connection
when modeling agents as finite automata, but her analysis is complex
(and holds only in the limit).

As we now show, the first-impression-matters bias can be easily
explained if we assume that there is a small cost for ``absorbing'' new
information. 
Consider the following simple game (which is very similar to the one
studied by Mullainathan \citeyear{M98} and Wilson \citeyear{W02}). 
The state of nature is a bit $b$ that is $1$ 
with probability $1/2$. 
For simplicity, we assume that the agent has no uncertainty about the
``goodness'' or output of a TM; the only uncertainty involves whether
$b$ is 0 or 1).
An agent receives as his type a sequence of
independent samples $s_1, s_2,\ldots, s_n$ where $s_i = b$ with
probability $\rho > 1/2$. The samples corresponds to signals the agents
receive about $b$.  
An agent is supposed to output a guess $b'$ for the bit $b$. If the
guess is correct, he receives $1-mc$ as utility, and $-mc$ otherwise,
where $m$ is the number of bits of the type he read, and $c$ is the cost
of reading a single bit ($c$ should be thought of the cost of
absorbing/interpreting information). 
It seems
reasonable to assume that $c>0$; signals usually require some effort to
decode (such as reading a newspaper article, or attentively watching a
movie). 
If $c>0$, it easily follows by the Chernoff bound (see \cite{AlSp04})
that after 
reading a certain (fixed) number of signals $s_1, \ldots, s_i$, the
agents will have a sufficiently good estimate of $b$ that the
marginal cost of reading one extra signal $s_{i+1}$ is higher than the
expected gain of finding out the value of $s_{i+1}$. 
That is, after processing a certain number of signals, agents will
eventually  disregard all future signals and base their output 
guess only on the initial sequence. 
We omit the straightforward details.
Essentially the same approach allows us to capture belief polarization.

Suppose for simplicity that two agents start out with slightly different
beliefs regarding the value of some random variable $X$ (think of $X$ as
representing something like ``O.J. Simpson is guilty''), and get the
same sequence $s_1, s_2, \ldots, s_n$  of evidence regarding the value
of $X$.  (Thus, now the type consists of the initial belief, which can
for example be modeled as a probability or a sequence of evidence
received earlier, and the new sequence of evidence.)  Both agents update
their beliefs by conditioning.  As before, 
there is a cost of processing a piece of evidence, so once an agent gets
sufficient evidence for either $X=0$ or $X=1$, he will stop processing
any further evidence.  If the initial evidence supports $X=0$, but the
later evidence supports $X=1$ even more strongly, the agent
that was initially inclined towards $X=0$ may raise his beliefs to be
above threshold, and thus stop
processing, believing that $X=0$, while the agent initially 
inclined towards $X=1$ will continue processing and eventually
believe that $X=1$.
}
\qed
\end{example}
As shown in \cite{HP10a}, we can also use this approach to explain  the 
\emph{status quo bias} (people are much more likely to stick with what
they already have) \cite{SZ88}.

\paragraph{Value of computational information and value of
conversation:} Having a computational model of decision making also
allows us to 
reconsider a standard notion from decision theory,
\emph{value of information}, and extend it in a natural way so as to
take computation into account. 
Value of information is meant to be a measure of how much an agent
should be  willing to 
pay to receive new information.  The idea is that, before receiving the
information, the agent has a probability on a set of relevant events and
chooses the action that maximizes his expected utility, given that
probability.  If he receives new information, he can update his
probabilities (by conditioning on the information) and again choose the
action that maximizes his expected utility.  The difference between the
expected utility before and after receiving the information is the value
of the information.

In many cases, an agent seems to be receiving valuable information that
is not about what seem to be the relevant events.  This means that we
cannot do a value of information calculation, at least not in the
obvious way.  

For example, suppose that the agent is interested in
learning a secret, which we assume for simplicity is a number between 1
and 1000.  A priori, suppose that the agent takes each number to be
equally likely, and so has probability $.001$.   
Learning the secret has utility, say, \$1,000,000; not learning it has
utility 0.  The number is locked in
a safe, whose combination is a 40-digit binary number.  Intuitively,
learning the first 20 digits of the safe's combination gives the agent some
valuable information.  But this is not captured when we do a standard
value-of-information calculation; learning this information has no
impact at all on the agent's beliefs regarding the secret.

Although this example is clearly contrived, there are many far more
realistic situations where people are clearly 
willing to pay for information to improve computation.  For example,
companies pay to learn about a manufacturing process that will speed up
production; people buy books on speedreading; 
and faster algorithms for search are clearly considered valuable.

Once we bring computation into decision making, the standard definition
of value of information can be applied to show that there is indeed a
value to learning the first 20 digits of the combination, and to buying
a more powerful computer; expected utility can increase.  (See
\cite{HP10} for details.)
 
But now we can define a new notion: \emph{value of conversation}.  The
value of information considers the impact of learning the value of a
random variable; by taking computation into account, we can extend this to
consider the impact of learning a better algorithm.  We can further
extend to consider the impact of having a conversation.  The point of a
conversation is that it allows the agent to ask questions based on
history.  For example, if the agent is trying to guess a number  chosen
uniformly at random between 1 and 100, and receives utility of 100 if he
guesses it right, having a conversation with a helpful TM that will
correctly answer seven yes/no questions is quite valuable:  as is well
known, with seven questions, the agent can completely determine the
number using binary search.  The computational 
framework
allows us to make this
intuition precise.  Again, we encourage the reader to consult
\cite{HP10} for further details.

\section{Modeling people as rational finite automata}\label{sec:aaai12}
We now consider the second approach discussed in the introduction, that
of modeling the fact that an agent can do only bounded computation.  Perhaps
the first to do this in the context of decision theory was Wilson
\citeyear{W02}, whose work we briefly mentioned earlier.
Recall that Wilson considers a decision problem where an agent needs to
make a single decision.  Nature is in one of two possible
states, $G$ (good) and $B$ (bad), which does not change over time; the
agent's payoff depends on the action she chooses and the state of nature.  
The agent gets one of $k$ signals, which are correlated with nature's state;
signal $i$ has probability $p_i^G$ of appearing when the state
of nature is $G$, and probability $p_i^B$ of appearing when the state is
$B$.  
We assume that the agent gets exactly one signal at each time step, so
that $\sum_{i=1}^k p_i^G = \sum_{i=1}^k p_i^B = 1$.
This is a quite standard situation.  For example, an agent may be on a
jury, trying to decide guilt or innocence, or a scientist trying to
determine the truth of a theory.

Clearly, the agent should try to learn nature's state, so as to make an
optimal decision.  
With no computational bounds, an agent should just keep track of all the
evidence it has seen (i.e., the number of signals of each type), in
order to make the best possible decision.  However, a finite automaton
cannot do this.  
Wilson characterizes the optimal $N$-state 
automaton for making a decision in this  setting, under the assumption
that $\eta$ (the probability that the agent has to make the decision in any
given round) is small.  Specifically, she shows that an optimal
$N$-state automaton  ignores all but two signals (the ``best'' signal
for each of nature's states).  The automaton's states can be laid out
``linearly'', as states 0, \ldots, $N-1$, and the automaton moves
left (with some probability) only if it gets a strong signal for state
$G$ (and it is not in state 0),
and moves right (with some probability) only if it gets a strong signal
for state $B$ (and is not in state $N-1$).  Roughly speaking, the lower
the  current state of the  
automaton, the more likely from the automaton's viewpoint that nature's
state is $G$. 

The probability of moving left of right, conditional on receiving the
appropriate signal, 
may vary from state to state.  In particular, if the automaton is in
state $0$, the probability 
that it moves right is very low.  Intuitively, in state 0, the automaton
is ``convinced'' that nature's state is $G$, and it is very reluctant to
give up on that belief.  Similarly, if the automaton is in state $N-1$,
the probability that it will move left is very low.  

Wilson argues that these
results can be used to explain observed biases in information 
processing, such as belief polarization.  For suppose that the true
state of nature is $B$.  Consider two 5-state automata, $A_1$ and
$A_2$.  Suppose that automaton $A_1$ starts in state 1,
while automaton $A_2$ starts in state 2.  (We can think of the starting 
state as reflecting some initial bias, for example.)  They both receive
the same information.  Initially, the information is biased towards $G$,
so both automata move left; $A_1$ moves to state 0, and $A_2$ moves to
state 1.  Now the evidence starts to shift towards the true state of the
world, $B$.  But since it is harder to ``escape'' from state 0, $A_1$
stays in state 0, while $A_2$ moves to state 4.  Thus, when the automata
are called upon to decide, $A_1$ makes the decision appropriate for $B$,
while $A_2$ makes the decision appropriate for $G$.  
A similar argument
shows how this approach can be used to explain the 
first-impression bias.
The key point is that the order in which evidence is received can make a
big difference to an optimal finite automaton, although it should make
no difference to an unbounded agent.

Wilson's model assumes that the state of nature never changes.
In recent work, we consider what happens if
we allow nature's state to change \cite{HPS12}.
We consider a model that is intended to
capture the most significant features of such a dynamic situation.  As in
Wilson's model, we allow nature to be in one of a number of different
states, and assume that the agent gets signals 
correlated with nature's state.  But now we allow nature's state to
change, although we assume that the probability of a change is low.
(Without this assumption, the signals are not of great interest.)  

For definiteness, assume that nature is in one of two states, which we
again denote $G$ and $B$. 
Let $\pi$ be the probability of transitioning from $B$ to $G$ or from $B$
to $G$ in any given round.  Thus, we assume for simplicity that these
probabilities are history-independent, and the same for each of the two
transitions.  (Allowing different probabilities in each direction does
not impact the results.)
The agent has two possible actions $S$ (safe) and $R$ (risky). If he
plays $S$, he gets a payoff of 0; if he plays $R$ he gets a 
payoff $x_G > 0$ when nature's state is $G$, and a payoff $x_B < 0$ when
nature's state is $B$.  
The agent does not learn his payoff, but, as in Wilson's model, gets one of
$k$ signals, whose probability is correlated with nature's state.
However, unlike Wilson's model, the agent gets a signal only if he plays the
risky action $R$; he does not get a signal if he plays the safe 
action $S$.
We denote this setting $S[p_1^G, p_1^B,\ldots, p_k^G,p_k^B, x_G,x_B]$. 
A setting is \emph{nontrivial} if there exists some signal
$i$ such that $p_i^B \ne p_i^G$.  If a setting is trivial, then no
signal enables the agent to distinguish whether nature is in state $G$
or $B$; the agent does not learn anything from the signals.  
For a given setting $S[p_1^G, p_1^B,\ldots, p_k^G,p_k^B, x_G,x_B]$, we
are interested in finding an automaton $A$ that has high average utility
when we let the number of rounds go to infinity.

Unlike Wilson, we were not able to find a characterization of the optimal
$N$-state automaton.  However, we were able to find a family of quite
simple automata that do very well in practice and, in the limit,
approach the optimal payoff.  We denote a typical member of this family
$A[N,p_{\exp},\Pos,\Neg,r_u,r_d]$.  
The automaton $A[N,p_{\exp},\Pos,\Neg,r_u,r_d]$ has $N+1$ states, again
denoted $0, \ldots, N$. State $0$ is dedicated to playing $S$. In all
other states 
$R$ is played. As in Wilson's optimal automaton,
only ``strong'' signals are considered; the rest are ignored.
More precisely, the $k$ signals are partitioned into three sets, $\Pos$ (for
``positive''), $\Neg$ (for ``negative''), and $I$ (for ``ignore'' or
``indifferent''),  with $\Pos$ and $\Neg$ nonempty. The
signals in $\Pos$ make it likely that nature's state is $G$, and the
signals in $\Neg$ make it likely that the state of nature is $B$. The
agent chooses to ignore the signals in $I$; they are viewed as not being
sufficiently informative as to the true state of nature.
(Note that $I$ is determined by $\Pos$ and $\Neg$.)

In each round while in state $0$, the agent moves to state
$1$ with probability $p_{\exp}$. 
In a state $i > 0$, if the agent receives a signal in $\Pos$, the agent
moves to $i+1$ with probability $r_u$ (unless he is already in state
$N$, in which case he stays in state $N$ if he receives a signal in $\Pos$); 
thus, we can think of $r_u$ as the probability that the agent moves up if
he gets a positive signal. If the agent receives a signal in $\Neg$, the
agent moves to state $i-1$ with probability $r_d$ (so $r_d$ is the
probability of moving down if he gets a signal in $\Neg$); if he
receives a signal in $I$, the agent does not change states.   Clearly,
this automaton is easy for a human to implement (at least, if it does
not have too many states).  Because the state of nature can change, it
is clearly not optimal to make the states $0$ and $N$ ``sticky''.  In
particular, an optimal agent has to be able to react reasonably quickly to a
change from $G$ to $B$, so as to recognize that he should play $S$.

Erev, Ert, and Roth \citeyear{erev2010entry} describe contests that
attempt to test various models of human decision making under uncertain
conditions. In their scenarios, people were given a 
choice between making a safe move (that had a guaranteed constant
payoff) and a ``risky'' move (which had a payoff
that changed according to an unobserved action of other players), in
the spirit of our $S$ and $R$ moves.
They challenged researchers to present models that would predict behavior in 
these settings.  The model that did best was called \emph{BI-Saw}
(bounded memory, inertia, sampling and  weighting) model, suggested by
Chen et al.~\citeyear{BI-SAW}, itself a refinement of a model called
\emph{I-Saw} suggested by Erev, Ert, and Roth \citeyear{erev2010entry}.
This model has three types of response mode:
\emph{exploration}, \emph{exploitation}, and \emph{inertia}.
An I-Saw agent proceeds as follows.
The agent tosses a coin.  If it lands heads, the agent plays
the action other than the one he played in the previous step
(\emph{exploration}); if it lands tails, he continues to do what he
did in the previous step (\emph{inertia}), unless
the signal received in the previous round
crosses a probabilistic ``surprise" trigger (the lower
 the probability of the signal to be observed in the current state, the
 more likely the trigger is to be crossed); 
if the surprise trigger is crossed, then the agent plays the
action with the best estimated 
subjective value, based on 
some sampling of the observations seen so far (\emph{exploitation}).
The major refinement suggested by BI-Saw involves adding a bounded
memory assumption, whose  main effect is a greater reliance on a small
sample of past observations. 

The suggested family of automata incorporates all three behavior modes
described by the I-Saw model. When 
the automaton is in state $0$, the agent \emph{explores} with constant
probability by moving to state $1$. 
In state $i > 0$, the agent continues to do what he did before (in
particular, he stays in state $i$) unless he gets a ``meaningful'' signal
(one in $\Neg$ or $\Pos$),
and even then he reacts only with some probability, so we have
\emph{inertia}-like behavior.
If he does react, then he \emph{exploits} the information that he has,
which is  
carried by his current state;
that is, he performs the action most appropriate according to his state.
The state can be viewed as representing a sample of the last few signals
(each state represents remembering seeing one more ``good" signal), as
in the BI-Saw model.   Thus, our family of automata can be viewed as an
implementation of the BI-Saw model using small, simple finite automata.

These automata do quite well, both theoretically and in practice.  
Note that even if the agent had an oracle that told him 
exactly what nature's state would be at every round, if he performs optimally, 
he can get only $x_G$ in the rounds when nature is in state $G$, and $0$ when
it is in state $B$. In expectation, nature is in state $G$ only half the
time, so the optimal expected payoff is $x_G/2$.  

The following result shows that 
if $\pi$ goes to 0 sufficiently quickly, then the agent can achieve
arbitrarily close to the theoretical optimum using an automaton of the form
$A[N,p_{\exp},\Pos,\Neg,r_u,r_d]$, even without the benefit
of an oracle, by choosing the parameters appropriately.  
Let $E_{\pi} [A]$ denote the expected average utility of using the
automaton $A$ if 
the state of natures changes with probability $\pi$.
\begin{theorem} \label{optimalAtLimit} {\rm \cite{HPS12}}
Let $\Pi$ and $P_{\exp}$ be functions from $\IN$ to $(0,1]$ such
that $\lim_{n \to \infty} n \Pi(n)  = \lim_{n \to \infty}
\Pi(n)/P_{\exp}(n) = 0$.  Then 
for all settings $S[p_1^G, p_1^B,\ldots, p_k^G,p_k^B, x_G,x_B]$, 
there exists a partition $\Pos,\Neg,I$ of the signals,
and constants $r_d$ and $r_u$ 
such that 
$$\lim_{N\to \infty}E_{\pi(N)}
[A[N,P_{exp}(N),\Pos,\Neg,r_u,r_d]]=\frac{x_G}{2}.$$     
\end{theorem}

\commentout{
Note that in Theorem~\ref{optimalAtLimit}, 
$\pi(N)$ goes to 0 as $N$ goes to infinity.
This requirement is necessary, as the next result shows; for fixed
$\pi$, we can't get too close to the optimal no matter what automaton we
use (in fact, the theorem holds even if the agent uses an
arbitrary strategy, not necessarily representable as an automaton).
\begin{theorem} {\rm \cite{HPS12}}
For all fixed $0<\pi\leq 0.5$ and all automata $A$,
we have $E_\pi[A] \le (1-\pi)x_G/2 + \pi x_B/2$.
\end{theorem}
}

While Theorem~\ref{optimalAtLimit} gives theoretical support to the
claim that automata of the form $A[N,p_{\exp},\Pos,\Neg,r_u,r_d]$ are
reasonable choices for a resource-bounded agent, it is also interesting
to see how they do in practice, for relatively small values of $N$.
The experimental evidence \cite{HPS12} suggests they will do well.  For
example,  suppose for definiteness that 
$\pi=0.001$, 
$x_G = 1$, $x_B = -1$ and
there are four signals, $1,
\ldots, 4$, which 
have probabilities $0.4$, $0.3$, $0.2$, and $0.1$, respectively, when
the state of nature is $G$, and probabilities $0.1$, $0.2$, $0.3$, and
$0.4$, respectively, when the state of nature is bad.   Further suppose
that we take  signal $1$ to be the ``good'' signal (i.e., we take 
$\Pos =\{1\}$), take signal $4$ to be the
``bad'' signal (i.e., we take $\Neg = \{4\}$),
and take $r_u = r_d = 1$. 
Experiments showed that using the optimal value of $p_{exp}$ (which
is dependent on the number of states),  
an automaton with $5$ states already gets an
expected payoff of more than $0.4$; even with
2 states, it gets an  
expected payoff of more than $0.15$.  Recall
that even with access to an 
oracle that reveals when nature changes state, the best the agent can
hope to get is $0.5$.  On the other hand, an agent that just plays
randomly, or always plays $G$ or always plays $B$, will get 0.  
These results are quite robust.  For example, for $N\geq 5$ the payoff
does not vary 
much if we just use the optimum $p_{\exp}$ for $N=5$, 
instead of choosing the optimum value of $p_{\exp}$ for each value of
$N$. 
The bottom line here is that, by thinking in terms of the algorithms
used by bounded agents, actions that may seem irrational can be viewed
as quite rational responses, given resource limitations.

\section{Discussion and Conclusion}
We have discussed two approaches for capturing resource-bounded
agents.  The first allows them to choose a TM to play for them, but
charges them for the ``complexity'' of the choice; the second 
models agents as finite automata, and captures resource-boundedness by
restricting the number of states of the automaton.  In both cases,
agents are assumed to maximize utility.
Both approaches can be used to show that some
systematic deviations from rationality (e.g., belief polarization) can
be viewed as the result of resource-bounded agents making quite rational
choices. 
We believe that the general idea of not viewing
behavior as ``irrational'', but rather the outcome
of resource-bounded agents making rational choices, will 
turn out to be useful for explaining other systematic biases in decision
making and, more generally, behavior in decision problems.
We would encourage work to find appropriate cost models, 
and simple, easy-to-implement strategies 
with low computational cost that perform well in real
scenarios.

Note that the two approaches are closely related.  For example, we can
easily come up with a cost model and class of TMs that result in agents
choosing a particular automaton with $N$ states to play for them, simply
by charging appropriately for the number of states.  Which approach is
used in a particular analysis depends on whether the interesting feature
is the choice of the complexity function (which can presumably be tested
experimentally) or specific details of the algorithm.  We are currently
exploring both approaches in the context of the behavior of agents in
financial markets.  In particular, we are looking for simple,
easy-to-implement strategies that will explain human behavior in this
setting. 

\bibliographystyle{plain}
\bibliography{lior,z,joe}
\end{document}